\documentclass[12pt]{article}
\usepackage{amssymb}

\usepackage{graphicx}
\usepackage{amsmath}

\input{tcilatex}

\begin{document}

\title{WAVE MECHANICS AND GENERAL RELATIVITY: A RAPPROCHEMENT}
\author{Paul S. Wesson \\
Department of Physics, University of Waterloo,\\
Waterloo, Ontario \ N2L 3G1, Canada}
\maketitle

\underline{{\Large Abstract}}{\Large :}

Using exact solutions, we show that it is in principle possible to regard
waves and particles as representations of the same underlying geometry,
thereby resolving the problem of wave-particle duality.

\bigskip

\underline {Scheduled to appear in Vol. 38, May 2006, of J. Gen. Rel. and Grav.}

\bigskip

\bigskip

Correspondence: Mail to address above, fax=(519)746-8115

\newpage

\section{\protect\underline{Introduction}}

Wave-particle duality is commonly presented as a conceptual conflict between
quantum and classical mechanics. \ The archetypal example is the
double-slit experiment, where electrons as discrete particles pass through a
pair of apertures and show wave-like interference patterns. \ However,
particles and waves can both be given geometrical descriptions, which raises
the possibility that these behaviours are merely different representations
of the same underlying geometry. \ We will give a brief discussion involving
exact solutions of extended geometry, to show that particles and waves may
be the same thing viewed in different ways.

Certain technical results will be needed below. \ (Those readers more
interested in results than method may like to proceed to section 2.) \ The
basic idea is that waves and particles are different coordinate
representations of the same geometry, or isometries [1-5]. \ Even in special
relativity, which frequently uses as a basis four-dimensional Minkowski
space $\left( M_{4}\right) ,$ we can if we so wish change the form of the
metric by a change of coordinates (or gauge). \ Thus, $M_{4}$ is actually
isometric to the Milne universe, which is often presented as a
Friedmann-Robertson-Walker (FRW) model with negative 3D or spatial curvature
in general relativity [4]. \ While the metrics may look different, their
equivalence is shown by the fact that in both cases the density and pressure
of matter are zero as determined by the field equations. \ The latter in 4D
read $R_{\alpha \beta }-Rg_{\alpha \beta }/ 2+\Lambda g_{\alpha \beta
}=8\pi T_{\alpha \beta }(\alpha ,\beta =0,123$ for time and space, where the
speed of light and the constant of gravity have been set to unity). \ Here $%
R_{\alpha \beta }$ is the Ricci tensor, $R$ is the Ricci scalar, $g_{\alpha
\beta }$ is the metric tensor, $\Lambda $ is the cosmological constant and $%
T_{\alpha \beta }$ is the energy-momentum tensor. \ While certain wave-like
solutions of the latter equations are known [3], none has the properties of
the deBroglie waves which are commonly used to describe the energy $\left(
E\right) $ and spatial momenta $p_{123}$ of particles in wave mechanics. \
Symbolically, these have wavelengths $\lambda ^{0}=h/ E,\;\lambda
^{1}=h/ p_{1}$ etc., where $h$ is Planck's constant (which may also be
set to unity). \ However, solutions of the field equations \underline{are}
known with deBroglie-like waves in dimensionally-extended gravity [5-7]. \
The latter is fundamentally Einstein's theory of general relativity,
extended to $N\left( >4\right) D$, in order to unify gravity with the
interactions of particle physics. \ The basic extension is to $N=5$, where
Campbell's theorem ensures that any solution of the 5D field equations in
vacuum is also a solution of the 4D field equations with matter [1, 8]. \
That is, we can always recover a solution of the 4D equations noted above
from the 5D equations, which in terms of the extended Ricci tensor are just $%
R_{AB}=0\left( A,B=0,123,4\right) $. \ There are many exact solutions known
of these equations, whereby the extended version is known to agree with
observations, both in regard to the solar system [5, 9] and cosmology [5,
10]. \ Several are relevant to the present project [11-14]. \ For example,
the Billyard solution [14] has a metric coefficient for the 3D or spatial
part which naturally represents the 3D or momentum component of a deBroglie
wave [15]. \ It is a remarkable solution, in that it is not only Ricci-flat $%
\left( R_{AB}=0\right) $ but also Riemann-flat $\left( R_{ABCD}=0\right) $.
\ That is, it represents a flat 5D space, which by virtue of Campbell's
theorem satisfies Einstein's 4D equations, and has a 3D deBroglie wave. \
However, it is deficient in some respects as regards the present project,
notably in that it has a signature $\left( +---+\right) $ which is at
variance with the one $\left( +----\right) $ indicated by particle physics.
\ The latter subject is constrained by Lorentz invariance and experiments
related to this. \ There is a relation between the energy $E$, 3-momentum $p$
and rest mass $m$ of a particle, which is regarded as standard because it is
closely obeyed in experiments (see ref. 16 for a review). \ Namely,%
\begin{equation}
E^{2}=p^{2}+m^{2}\;\;\;\;\;.
\end{equation}%
This is a strong constraint on any attempt to construct a geometric relation
between the particle and wave descriptions of matter. \ From the viewpoint
of a theory like general relativity, (1) is perhaps not surprising, in that
it can be understood as a consequence of multiplying a constant $m$ onto the
conventional condition for normalizing the 4-velocities, viz\ $u^{\alpha
}u_{\alpha }=1$. \ (Here, $u^{\alpha }\equiv dx^{\alpha }/ds$ where the 4D
coordinates $x^{\alpha }$ are related to the proper time $s$ and the metric
tensor $g_{\alpha \beta }$ via $ds^{2}=g_{\alpha \beta }dx^{\alpha
}dx^{\beta }$.) \ From the viewpoint of a dimensionally-extended theory, (1)
is also not so surprising, in that it follows for a wide range of metrics. \
The latter involve two aspects. First, the coordinates should be
``canonical'' in 5D, which means that the interval can be written as $%
dS^{2}=\left( l/ L\right) ^{2}ds^{2}-dl^{2}$ where $x^{4}\equiv l$, so
that the extra coordinate plays the role of particle mass and the Weak
Equivalence Principle is obeyed [17, 18]. \ Second, the paths of particles
(or waves) should be null, so that a photon-like object in 5D appears as a
massive object in 4D [19-21]. \ This latter condition enables us to cast our
project into a new form: we are asking if there is a photon-like solution in
5D, which in 4D can be interpreted as being a massive particle or
equivalently a deBroglie wave.

The technical results noted in the preceding paragraph may appear to be very
restrictive as regards a possible resolution of the apparent dichotomy
between wave mechanics and classical mechanics. \ However, wave-particle
duality is a generic feature of matter, as shown by experiments more
sophisticated than the old double-slit one for electrons, such as studies of
the interference in a gravitational field of neutrons [22, 23]. \ It is
therefore to be expected that if there is a geometric explanation in terms
of isometries, that it also will be generic in some sense. \ This will turn
out to be the case. \ Thus while there are numerous coordinate frames which
are useful for our studies, it transpires that they share the property of
describing a 5D manifold that is flat [24]. \ We will present two exact
solutions which represent deBroglie waves but share this property in section
2. \ The inference is that particles and waves in 4D are isometries of flat
5D space.

\section{\protect\underline{Exact Wave-Particle Solutions}}

In this section, we will consider flat manifolds of various
dimensionalities, with a view to showing that a 4D deBroglie wave which
describes energy and momentum is isometric to a flat 5D space. \ The
notation is the same as that introduced above, and standard.

2D manifolds, like that which describes the surface of the Earth, are locally
flat. \ A brief but instructive account of their isometries is given by Rindler
(1977, p.114; a manifold of any N is approximately flat in a small enough region, and
changes of coordinates that qualify as isometries should strictly speaking preserve
the signature.) \ Consider, as an example, the line
element $ds^{2}=dt^{2}-t^{2}dx^{2}$. \ Then the coordinate transformation $%
t\rightarrow e^{i\omega t}/ i\omega $, $x\rightarrow e^{i\kappa x}$
causes the metric to read $ds^{2}=e^{2i\omega t}dt^{2}-e^{2i\left( \omega
t+\kappa x\right) }dx^{2}$, where $\omega $\ is a frequency, $\kappa $ is a
wave-number and the phase velocity $\omega /\kappa $ has been set to unity.
\ It is clear from this toy example that a metric which describes a
freely-moving particle (the proper distance is proportional to the time) is
equivalent to one which describes a freely-propogating wave.
\ For the particle, we can define its energy and
momentum via $E\equiv m\left( dt/ ds\right) $ and $p\equiv m\left(
dx/ ds\right) $. \ For the wave, $\widetilde{E}\equiv m\,e^{i\omega
t}\left( dt/ ds\right) $ and $\widetilde{p}\equiv m$ $e^{i\left(
\omega t+\kappa x\right) }\left( dx/ ds\right) $. \ In both cases, the
mass $m$ of a test particle has to be introduced \underline{ad hoc}, a
shortcoming which will be addressed below. \ The standard energy condition
(1), in the form $m^{2}=E^{2}-p^{2}$, is recovered if the signature is $%
\left( +-\right) $. \ If on the other hand we have a Euclidean signature of
the kind used in certain approaches to quantum gravity, it is instructive in
the 2D case to consider the isometry $ds^{2}=x^{2}dt^{2}+t^{2}dx^{2}$. \ The
transformation $t\rightarrow e^{i\omega t}/ i\omega $, $x\rightarrow
e^{i\kappa x}/ i\kappa $ causes this to read $ds^{2}=-\left( 1/
\kappa \right) ^{2}\,\,e^{2i\left( \omega t+\kappa x\right) }$ $\left(
dt^{2}+dx^{2}\right) $, after the absorption of a phase velocity as above. \
Thus a particle metric becomes one with a conformal factor which resembles a
wave function.

3D manifolds add little to what has been discussed above. \ It is well known
that in this case the Ricci and Riemann-Christoffel tensors can be written
as functions of each other, so the field equations bring us automatically to
a flat manifold as before.

4D manifolds which are isotropic and homogeneous, but non-static, lead us to
consider the FRW metrics. \ These have line elements given by%
\begin{equation}
ds^{2}=dt^{2}-\frac{R^{2}\left( t\right) }{\left( 1+k\;r^{2}/ 4\right)
^{2}}\left( dx^{2}+dy^{2}+dz^{2}\right) \;\;\;\;\;,
\end{equation}%
where $R\left( t\right) $ is the scale factor and $k=\pm $1, 0 defines the
3D curvature. \ (This should not be confused with the wave number.) \ In the
ideal case where the density and pressure of matter are zero, a test
particle moves away from a local origin with a proper distance proportional
to the time. \ (I.e., $R=t$ above where the spatial coordinates $xyz$ and $%
r\equiv \sqrt{x^{2}+y^{2}+z^{2}}$ are comoving and dimensionless.) \ This
specifies the Milne model, which by the field equations requires $k=-1$. \
(One way think of this as a situation where the kinetic energy is balanced
by the gravitational energy of a negatively-curved 3D space.) \ As noted in
section 1, (2) with $R=t$ and $k=-1$ is isometric to $M_{4}$ [ref.4, p.205].
\ Indeed, the Milne model is merely a convenient non-static representation
of flat 4D space. \ In the local limit where $\left| r^{2}/ 4\right|
\ll 1$, the t-behaviour of the 3D sections of (2) allows us to specify
a wave via the same kind of coordinate transformation used in the 2D case. \
We eschew the details of this, since the same physics is contained in more
satisfactory form if the dimensionality is extended.

5D manifolds which are canonical [5, 24] have remarkably simple dynamics. \
And since Campbell's theorem [1, 8] ensures that any Ricci-flat 5D solution
has an Einstein 4D analog, it is natural to focus on the 5D version of the
Milne model discussed in the preceding paragraph. \ Consider therefore the
5D line element 
\begin{equation}
dS^{2}=\left( \frac{l}{L}\right) ^{2}dt^{2}-\left[ l\,\text{sinh}\left( 
\frac{t}{L}\right) \right] ^{2}d\sigma ^{2}-dl^{2}\;\;\;\;\;.
\end{equation}%
Here $l$ is the extra coordinate, and $L$ is a constant length which we will
see below is related inversely to the cosmological constant $\Lambda $. \ \
The 3-space is the same as that above, namely $d\sigma ^{2}=\left(
dx^{2}+dy^{2}+dz^{2}\right) $ $\left( 1+kr^{2}/ 4\right) ^{-2}$ with $%
k=-1$. \ That the time-dependence of the 3-space in (3) is different from
that in (2) is attributable to the fact that we are using the 4D parameters (%
$t$, $xyz$ or equivalently the 4D proper time $s$) to describe the motion in
a 5D metric (whose proper time is $S\neq s$: see note\ 24). \ However, the
local situation for the 5D case (3) is close to that for the 4D case (2). \
To see this, we note that for laboratory situations $t/ L\ll 1$ in
(3), so it reads 
\begin{equation}
dS^{2}\simeq \left( \frac{l}{L}\right) ^{2}dt^{2}-\left( \frac{lt}{L}\right)
^{2}d\sigma ^{2}-dl^{2}\;\;\;\;\;.
\end{equation}%
This is of canonical form, namely $dS^{2}=\left( l/ L\right)
^{2}ds^{2}-dl^{2}$ [5, 8, 14, 15, 18, 20]. \ For such metrics, the reduction
of the 5D field equations to the 4D Einstein equations identifies the length 
$L$ via $\Lambda =3/ L^{2}$ (see e.g. ref. 5, p. 159). \ Such metrics
effectively describe momentum manifolds rather than coordinate manifolds,
since the identification of $l$ with $m$ defines the conventional action of
particle physics $\left( \int mds\right) $, and ensures agreement with the
Weak Equivalence Principle [18]. \ More importantly for present purposes,
the metric (3) from which (4) is derived satisfies not only $R_{AB}=0$ but
also $R_{ABCD}=0$. \ This may be confirmed either by algebra or a fast
computer package such as GRTensor. \ Since the 5D manifold is flat, the
appropriate condition for the path of a particle in it is $dS=0$ [19, 20]. \
With this condition, any canonical metric results in the constraint $L\left(
dl/ ds\right) =\pm l$. \ Let us use this constraint with (4), where we
multiply it by $L^{2}$ and divide it by $ds^{2}$. \ The result is 
\begin{equation}
0\simeq l^{2}\left( \frac{dt}{ds}\right) ^{2}-\left( l\,\,t\right) ^{2}\left[
\left( \frac{dx}{ds}\right) ^{2}+\left( \frac{dy}{ds}\right) ^{2}+\left( 
\frac{dz}{ds}\right) ^{2}\right] -l^{2}\ \ \;\ \ .
\end{equation}%
This with the identification $l=m$ (see above) and the recollection that
proper distances are defined by $\int tdx$ etc., simply reproduces the
standard condition (1), in the form $0=E^{2}-p^{2}-m^{2}$.

To convert the 5D metric (4) to a wave, we follow the lower-dimensional
examples noted before. \ Specifically, we change $t\rightarrow e^{i\omega
t}/ i\omega $, $x\rightarrow \exp \left( i\kappa _{x}x\right) $ etc.,
where $\omega $\ is a frequency and $\kappa _{x}$ etc. are wave numbers for
the $x,y,z$ directions. \ After setting the phase velocity to unity, (4)
then reads 
\begin{equation}
dS^{2}\simeq \left( \frac{l}{L}\right) ^{2}e^{2i\omega t}dt^{2}-\left( \frac{%
l}{L}\right) ^{2}\left\{ \exp \left[ 2i\left( \omega t+\kappa _{x}x\right) %
\right] dx^{2}+\text{etc}\right\} -dl^{2}\;\;\;.
\end{equation}%
This with the null condition causes the analog of (5) to read%
\begin{equation}
0\simeq \left\{ l\;e^{i\omega t}\frac{dt}{ds}\right\} ^{2}-\left\{ l\exp %
\left[ i\left( \omega t+\kappa _{x}x\right) \right] \frac{dx}{ds}\right\}
^{2}-\text{etc}-l^{2}\;\;\;\;\;.
\end{equation}%
We can again make the identification $l=m$ and define 
\begin{equation}
\widetilde{E}\equiv l\,e^{i\omega t}\frac{dt}{ds},\;\;\widetilde{p}\equiv
l\exp \left[ i\left( \omega t+\kappa _{x}x\right) \right] \frac{dx}{ds}\;\;%
\text{etc}.
\end{equation}%
Then (7) is equivalent to%
\begin{equation}
0\simeq \widetilde{E}^{2}-\widetilde{p}^{2}-m^{2\;\;\;\;\;.}
\end{equation}%
This is of course the wave analog of the standard relation (1) for a
particle.

Another example of a 5D wave-like metric is the Billyard solution, which
like (3) above satisfies $R_{AB}=0$ and $R_{ABCD}=0$ [14, 15, 24]. \ It may
be expressed in a form somewhat different from the original as 
\begin{equation}
dS^{2}=\left( \frac{l}{L}\right) ^{2}dt^{2}-\left( \frac{l}{L}\right)
^{2}\left\{ \exp \left[ 2i\left( \frac{t}{L}+\kappa _{x}x\right) \right]
dx^{2}+\text{etc}\right\} +dl^{2}\ \ \;\;\ \ \ .
\end{equation}%
This metric resembles (6), but now the frequency is constrained by the scale
of the geometry and the extra dimension is timelike. \ (It may be verified
that there is no solution for the opposite case.) \ The latter property
means that for null 5D geodesics we have $l=l_{0}e^{\pm is/ L}$ where $%
l_{0}$ is a constant, so the mass parameter is itself a wave which
oscillates around the hypersurface we call spacetime [21]. \ Such behaviour
can also occur in string theory [6, 7], and may or may note be realistic. \
For present purposes, we note that while (10) can describe a deBroglie wave
for the 3-momentum, it is not clear how to treat the energy, and the
signature is at variance with that implied by the standard particle relation
(1). \ This may seem strange, given the similarities between (6) and (10). \
However, it should be recalled that in trying to identify a 4D deBroglie
wave from a 5D metric, we are dealing with a quantity $Q=Q\left( x^{\alpha
},l\right) $ which is \underline{not} necessarily preserved under the group
of 5D coordinate transformations $x^{A}\rightarrow \overline{x}^{A}\left(
x^{B}\right) $ if the extra one $x^{4}=l$ is involved. \ This implies that
the exact 5D solutions (3) and (10) are \underline{not} equivalent in terms
of their 4D physics. \ Indeed, the reduction of the field equations from 5D
to 4D implies that the approximate form (4) of (3) has $\Lambda >0$, whereas
(10) has $\Lambda <0$, due to their different signatures [5]. \ This and
other aspects of these solutions should be investigated in future work. \ At
present, it appears that duality can best be described by (3), (4) for the
particle and (6) for the wave.

\section{\protect\underline{Conclusion}}

Wave-particle duality may be approached through a consideration of flat
manifolds of various dimensionalities. \ In the context of classical 4D
general relativity, the standard energy condition (1) of particle physics in
vacuum is consistent with the Milne model (2). \ This is an isometry of
Minkowski space $M_{4}$, and a coordinate transformation can be used to make
it wave-like. \ However, the concept of momentum is better handled by $M_{5}$%
, and we have examined an exact solution (3) in 5D which is not only
Ricci-flat $\left( R_{AB}=0\right) $ but also Riemann-flat $\left(
R_{ABCD}=0\right) $. \ The local limit of this solution is (4), which is
basically the Milne model embedded in a 5D momentum (as opposed to
coordinate) manifold. \ This describes a particle which obeys the standard
energy condition, but a coordinate transformation puts a wave on it as in
(6). \ Since the underlying manifold is flat, the natural condition on the
interval (action) is that it be zero, as in (7). \ Then obvious definitions
for the energy and spatial momentum (8) result in both quantities being
wave-like, and obeying a wave analog (9) of the particle energy condition. \
The solution (3), while it lends itself easily to both particle and wave
interpretations, deserves further study to see what other physics it may
imply. \ By contrast, the solution (10) which has been discussed in the
literature is already in wave form, but does not lend itself so readily to
an interpretation in terms of deBroglie waves. \ Our main conclusion, based
on the solution (3), is that particles and waves are isometries of flat 5D
space.

This technical result invites a philosophical discussion which would be
inappropriate here. \ However, some comments are in order about coordinates.
\ Physics should always be constructed in an $N$-dimensional space in a
manner which is covariant; but if that space is embedded in $\left(
N+1\right) $, and the extra coordinate enters in a significant way, the
physics in $ND$ will necessarily depend on the coordinates in $\left(
N+1\right) D$. \ Traditional Kaluza-Klein theory is a good example of this,
where the electromagnetic potentials (which are the cross-terms in the
extended metric tensor) can be included or decluded depending on how the 5
degrees of coordinate freedom for the line element are used. \ Even in
manifolds of fixed $N$, the physical interpretation of a solution can depend
on the choice of coordinates or gauge. \ The Minkowski and Milne cases in 4D
provide a good example of this, where the former describes a static
spacetime and the latter describes an expanding cosmology. \ (Certain
quantities are of course preserved, and in this case the density and
pressure of matter are zero in both interpretations.) \ Likewise, the
particle and wave descriptions for energy and momentum which we have
discussed above depend on a choice of coordinates. \ The waves are not
electromagnetic, and nor are they gravitational of the conventional type. \
For want of a better term, they can be called metric waves. \ They should
not be regarded as merely technical accidents. \ Physics has over a long
period given us large bodies of information which, because of the
experimental approaches involved, we describe as pertaining to particles and
waves. \ But it is really not surprising that these two physical phenomena
have a common mathematical base. \ We have simply argued that this common
base is geometrical, and that particles and waves are isometries.

Wave-particle duality has long been considered a paradox, but it may simply
be that particles and waves are the same thing viewed in geometrically
different ways.

\bigskip

\bigskip

\bigskip

\underline{{\Large Acknowledgements}}

The outline given here has benefited from discussions over time with various
people, including A. Billyard, T. Liko, W. Rindler and S. Werner. \ The work
was supported in part by NSERC.

\underline{{\LARGE References}}

\begin{enumerate}
\item Campbell, J. 1926. A Course on Differential Geometry (Oxford:
Clarendon Press).

\item Eisenhart, L.D. 1949. Riemannian Geometry (Princeton: Princeton
University Press).

\item Kramer, D., Stephani, H., Herlt, E., MacCallum, M., Schmutzer, E.
1980. \ Exact Solutions of Einstein's Field Equations (Cambridge: Cambridge
University Press).

\item Rindler, W. 1977. \ Essential Relativity (New York: Springer).

\item Wesson, P.S. 1999. \ Space, Time, Matter (Singapore: World Scientific).

\item Gubser, S.S., Lykken, J.D. (editors), 2004. \ Strings, Branes and
Extra Dimensions (Singapore: World Scientific).

\item Szabo, R.J., 2004. \ An Introduction to String Theory and D-Brane
Dynamics (Singapore: World Scientific).

\item Seahra, S.S., Wesson, P.S. 2003. \ Class. Quant. Grav. \underline{20},
1321.

\item Kalligas, D., Wesson, P.S., Everitt, C.W.F. 1995. \ Astrophys. J. 
\underline{439}, 548.

\item Ponce de Leon, J. 1988, Gen. Rel. Grav. \underline{20}, 539.

\item Davidson, A., Sonnenschein, J., Vozmediano, A.H. 1985. \ Phys. Rev. D 
\underline{32}, 1330.

\item McManus, D.J. 1994. \ J. Math. Phys. \underline{35}, 4889.

\item Abolghasem, G., Coley, A.A., McManus, D.J. 1996. \ J. Math. Phys. 
\underline{37}, 361.

\item Billyard, A., Wesson, P.S. 1996. \ Gen. Rel. Grav. \underline{28}, 129.

\item Wesson, P.S. 2003. \ Int. J. Mod. Phys. D \underline{12}, 1721.

\item Pospelov, M., Romalis, M. 2004. \ Phys. Today \underline{57} (7), 40.

\item Will, C.M. 1993. \ Theory and Experiment in Gravitational Physics
(Cambridge: Cambridge University Press).

\item Wesson, P.S. 2003. \ Gen. Rel. Grav. \underline{35}, 307.

\item Youm, D. 2001. \ Mod. Phys. Lett. A \underline{16}, 2371.

\item Seahra, S.S., Wesson, P.S. 2001. \ Gen Rel. Grav. \underline{33}, 1731.

\item Wesson, P.S., 2002. \ Phys. Lett. B \underline{538}, 159.

\item Colella, R., Overhauser, A.W., Werner, S.A., 1975. \ Phys. Rev. Lett. 
\underline{34}, 1472.

\item Rauch, H., Werner, S.A., 2000. \ Neutron Interferometry (Oxford:
Clarendon).

\item There are several useful coordinate transformations which relate flat
5D manifolds and curved 4D ones. \ For example, the metrics $%
dS^{2}=dT^{2}-d\sigma ^{2}-dL^{2}$ and $ds^{2}=l^{2}dt^{2}-d\sigma
^{2}-t^{2}dl^{2}$ are related by the transformation $T=t^{2}l^{2}/
4+\ln \left( t^{1/ 2}l^{-1/ 2}\right) $, $L=t^{2}l^{2}/
4-\ln \left( t^{1/ 2}l^{-1/ 2}\right) $. \ The ``standard'' 5D
cosmologies of Ponce de Leon [10] have metrics of the second-noted form, and
may by coordinate transformations be shown to be 5D flat. \ The full
transformations, including those for the spatial part, are given elsewhere
(ref. 5, p. 49). \ The Billyard wave [14] may similarly be shown to be a
coordinate-transformed version of de Sitter space, and flat in 5D. \ A
generic discussion of cosmological models which are flat in 5D is due to
McManus [12]. \ One of his solutions is a metric for a particle in a
manifold whose 3D part is curved, which effectively generalizes the Billyard
wave whose 3D part is flat. \ [See ref. 12, p. 4895, equation (30).] \ This
can be seen by changing the 4D coordinates as discussed in the main text,
which results in $dS^{2}=\left( l/ L\right) ^{2}dt^{2}-\left( l/
L\right) ^{2}\left( e^{it/ L}+ke^{-it/ L}\right) ^{2}\left[ \exp
\left( 2i\kappa _{x}x\right) dx^{2}+\text{etc}\right] +dl^{2}$. \ When the
curvature constant k is zero, this gives back the Billyard wave. \ Another
of the McManus solutions reproduces work by Davidson et al. \ [See ref. 11;
and also ref. 12, p. 4893, equation (19).] \ This is effectively a 5D
embedding of the 4D Milne model, and can be written as $dS^{2}=dt^{2}-t^{2}d%
\sigma ^{2}-dl^{2}$, where $d\sigma ^{2}\equiv \left( 1+kr^{2}/
4\right) ^{-2}$ with $k=-1$. \ The transformation $t\rightarrow l\,$sinh$%
\left( t/ L\right) $, $l\rightarrow l\ $cosh$\left( t/ L\right) $
causes the metric to read $dS^{2}=\left( l/ L\right) ^{2}dt^{2}-\left[
l\text{ sinh}\left( t/ L\right) \right] ^{2}d\sigma ^{2}-dl^{2}$. \
This is quoted as (3) of the main text, and its local approximation is (4).
\ The former has proper distances which vary as sinh$t$, whereas the latter
has proper distances which vary as $t$. \ The former is typical of motion in
flat 5D space, when the 4D proper time $s$ (as opposed to the 5D proper time 
$S$) is used as parameter [5, p. 169]. \ The latter is typical of motion in
flat 4D space, when the ordinary time $t$ is used as parameter [4, p. 205].
\ Both of the models used in the main text to illustrate the passage from
particle to wave use metrics which are canonical in form, and there is a
large literature on these. \ However, a more general class of metrics is
given by $dS^{2}=g_{\alpha \beta }\left( x^{\gamma },l\right) dx^{\alpha
}dx^{\beta }+\epsilon \Phi ^{2}\left( x^{\gamma },l\right) dl^{2}$, where $%
\epsilon =\pm 1$ and $\Phi $ is a scalar field. \ Einstein's 4D equations
are satisfied for this 5D metric if the effective or induced energy-momentum
tensor is given by%
\begin{eqnarray*}
8\pi T_{\alpha \beta } &=&\frac{\Phi _{,\alpha ;\beta }}{\Phi }-\frac{%
\epsilon }{2\Phi ^{2}}\left\{ \frac{\Phi ,_{4}g_{\alpha \beta ,4}}{\Phi }%
-g_{\alpha \beta ,44}+g^{\lambda \mu }g_{\alpha \lambda ,4}g_{\beta \mu
,4}\right.  \\
&&\;\;\;\;\;\;\;\;\;\left. -\frac{g^{\mu \nu }g_{\mu \nu ,4}g_{\alpha \beta
,4}}{2}+\frac{g_{\alpha \beta }}{4}\left[ g_{\;,4}^{\mu \nu }g_{\mu \nu
,4}+\left( g^{\mu \nu }g_{\mu \nu 4}\right) ^{2}\right] \right\}
\;\;\;\;\;\;\;\;\;.
\end{eqnarray*}%
Here a comma denotes the partial derivative and a semicolon denotes the 4D
covariant derivative. \ We have not discussed the matter which relates to
the exact solutions (3), (10) of the main text because it is merely vacuum
[5, 14]. \ But the matter properties of these and \ more complicated
solutions may be evaluated for any choice of coordinates by using the noted
expression.
\end{enumerate}

\end{document}